\documentclass[prd,twocolumn,aps,superscriptaddress]{revtex4}
\usepackage{feyn}
\usepackage{color}
\usepackage{hyperref}
\usepackage{graphicx}
\usepackage{amsmath}
\usepackage{bbold}
\usepackage{slashed}

\usepackage{amssymb}
\begin{document}

%


\newcommand{\Gb}{\pazocal{G}}
\newcommand{\be}{\begin{equation}}
\newcommand{\ee}{\end{equation}\noindent}
\newcommand{\bear}{\begin{eqnarray}}
\newcommand{\ear}{\end{eqnarray}\noindent}
\newcommand{\no}{\noindent}
\newcommand{\non}{\nonumber\\}

\def\veps#1{\varepsilon_{#1}}
\def\ddel{{}^\bullet\! \Delta}
\def\deld{\Delta^{\hskip -.5mm \bullet}}
\def\dddel{{}^{\bullet \bullet} \! \Delta}
\def\ddeld{{}^{\bullet}\! \Delta^{\hskip -.5mm \bullet}}
\def\deldd{\Delta^{\hskip -.5mm \bullet \bullet}}
\def\epsk#1#2{\varepsilon_{#1}\cdot k_{#2}}
\def\epseps#1#2{\varepsilon_{#1}\cdot\varepsilon_{#2}}
\def\eq#1{{eq. (\ref{#1})}}
\def\eqs#1#2{{eqs. (\ref{#1}) -- (\ref{#2})}}
\def\t#1{\tau_1}
\def\mn{{\mu\nu}}
\def\rs{{\rho\sigma}}
\newcommand{\Det}{{\rm Det}}
\def\Tr{{\rm Tr}\,}
\def\tr{{\rm tr}\,}
\def\sumij{\sum_{i<j}}
\def\e{\,{\rm e}}
\def\eps{\varepsilon}

\def\bddel{{}^\bullet\! {\underline\Delta}}
\def\bdeld{{\underline\Delta}^{\hskip -.5mm \bullet}}
\def\bdddel{{}^{\bullet \bullet} \! {\underline\Delta}}
\def\bddeld{{}^{\bullet}\! {\underline\Delta}^{\hskip -.5mm \bullet}}
\def\bdeldd{{\underline\Delta}^{\hskip -.5mm \bullet \bullet}}


\def\fr#1#2{\frac{#1}{#2}}
\def\half{\frac{1}{2}}
\def\freeexp{{\rm e}^{-\int_0^Td\tau {1\over 4}\dot x^2}}
\def\kinb{{1\over 4}\dot x^2}
\def\kinf{{1\over 2}\psi\dot\psi}
\def\expk{{\rm exp}\biggl[\,\sum_{i<j=1}^4 G_{Bij}p_i\cdot p_j\biggr]}
\def\expp{{\rm exp}\biggl[\,\sum_{i<j=1}^4 G_{Bij}p_i\cdot p_j\biggr]}
\def\expshort{{\e}^{\half G_{Bij}p_i\cdot p_j}}
\def\expabb{{\e}^{(\cdot )}}
\def\epseps#1#2{\varepsilon_{#1}\cdot \varepsilon_{#2}}
\def\epsk#1#2{\varepsilon_{#1}\cdot k_{#2}}
\def\epsr#1#2{r_{#2}\cdot\varepsilon_{#1}}
\def\rk#1#2{r_{#1}\cdot p_{#2}}
\def\kk#1#2{k_{#1}\cdot k_{#2}}
\def\G#1#2{G_{B#1#2}}
\def\Gp#1#2{{\dot G_{B#1#2}}}
\def\GF#1#2{G_{F#1#2}}
\def\Dab{{(x_a-x_b)}}
\def\Dsq{{({(x_a-x_b)}^2)}}
\def\PITD{{(4\pi T)}^{-{D\over 2}}}
\def\4piTD{{(4\pi T)}^{-{D\over 2}}}
\def\4piT4{{(4\pi T)}^{-2}}
\def\TintmD{{\dps\int_{0}^{\infty}}{dT\over T}\,e^{-m^2T}
    {(4\pi T)}^{-{D\over 2}}}
\def\Tintm4{{\dps\int_{0}^{\infty}}{dT\over T}\,e^{-m^2T}
    {(4\pi T)}^{-2}}
\def\Tintm{{\dps\int_{0}^{\infty}}{dT\over T}\,e^{-m^2T}}
\def\Tint{{\dps\int_{0}^{\infty}}{dT\over T}}
\def\np{n_{+}}
\def\nm{n_{-}}
\def\Np{N_{+}}
\def\Nm{N_{-}}
\def\ed{e^{(\cdot)}}
\def\t#1{\tau_{#1}}
\def\et#1#2{e^{({#1}\rightarrow{#2})}}
\def\ett#1#2#3#4{e^{({#1}\rightarrow{#2}, #3\rightarrow#4)}}
\newcommand{\slG}{{{\dot G}\!\!\!\! \raise.15ex\hbox {/}}}
\newcommand{\Gd}{{\dot G}}
\newcommand{\Gund}{{\underline{\dot G}}}
\newcommand{\Gdd}{{\ddot G}}
\def\GBd12{{\dot G}_{B12}}
\def\Dx{\dps\int{\cal D}x}
\def\Dy{\dps\int{\cal D}y}
\def\Dpsi{\dps\int{\cal D}\psi}
\def\dint#1{\int\!\!\!\!\!\int\limits_{\!\!#1}}
\def\ddtau{{d\over d\tau}}
\def\ie{\hbox{$\rm style{\int_1}$}}
\def\iz{\hbox{$\rm style{\int_2}$}}
\def\id{\hbox{$\rm style{\int_3}$}}
\def\ldop{\hbox{$\lbrace\mskip -4.5mu\mid$}}
\def\rdop{\hbox{$\mid\mskip -4.3mu\rbrace$}}
\def\bdel{{\underline\Delta}}
%
\newcommand{\1}{{\'\i}}
\def\dps{\displaystyle}
\def\sy{\scriptscriptstyle}
\def\sy{\scriptscriptstyle}

\def\del{\partial}
\def\deli{\partial_{\kappa}}
\def\delj{\partial_{\lambda}}
\def\delk{\partial_{\mu}}
\def\delij{\partial_{\kappa\lambda}}
\def\delik{\partial_{\kappa\mu}}
\def\deljk{\partial_{\lambda\mu}}
\def\delki{\partial_{\mu\kappa}}
\def\delkl{\partial_{\mu\nu}}
\def\delijk{\partial_{\kappa\lambda\mu}}
\def\deljkl{\partial_{\lambda\mu\nu}}
\def\delikl{\partial_{\kappa\mu\nu}}
\def\delijkl{\partial_{\kappa\lambda\mu\nu}}
\def\delijklm{\partial_{\kappa\lambda\mu\nu o}}
\def\O(#1){O($T^#1$)} 
\def\O2{O($T^2$)}
\def\O3{O($T^3$)}
\def\O4{O($T^4)}
\def\O5{O($T^5$)}
\def\dA{\partial^2}
\def\DA{\sqsubset\!\!\!\!\sqsupset}
\def\eins{  1\!{\rm l}  }
\def\a#1{\alpha_{#1}}
\def\b#1{\beta_{#1}}
\def\m#1{\mu_{#1}}
\def\n#1{\nu_{#1}}
\def\m#1{\mu_{#1}}
\def\n#1{\nu_{#1}}
\def\a{\alpha}
\def\b{\beta}
\def\m{\mu}
\def\n{\nu}
\def\s{\sigma}
\def\r{\rho}
\def\e{{\rm e}}
\def\z{\zeta}
\def\vareps{\varepsilon}

\def\gF{\gamma_{\mathcal{F}}}
\def\gG{\gamma_{\mathcal{G}}}
\def\gFF{\gamma_{\mathcal{F}\mathcal{F}}}
\def\gGG{\gamma_{\mathcal{G}\mathcal{G}}}
\def\gFG{\gamma_{\mathcal{F}\mathcal{G}}}

\newcommand{\Vka}{V_{\kappa}}
\newcommand{\Vla}{V_{\lambda}}
\newcommand{\Vmu}{V_{\mu}}
\newcommand{\Vnu}{V_{\nu}}
\newcommand{\Vro}{V_{\rho}}
\newcommand{\Vkala}{V_{\kappa\lambda}}
\newcommand{\Vkamu}{V_{\kappa\mu}}
\newcommand{\Vkanu}{V_{\kappa\nu}}
\newcommand{\Vlamu}{V_{\lambda\mu}}
\newcommand{\Vlanu}{V_{\lambda\nu}}
\newcommand{\Vlaka}{V_{\lambda\kappa}}
\newcommand{\Vmunu}{V_{\mu\nu}}
\newcommand{\Vmuka}{V_{\mu\kappa}}
\newcommand{\Vnuro}{V_{\nu\rho}}
\newcommand{\Vkalamu}{V_{\kappa\lambda\mu}}
\newcommand{\Vkalanu}{V_{\kappa\lambda\nu}}
\newcommand{\Vkalaro}{V_{\kappa\lambda\rho}}
\newcommand{\Vkamunu}{V_{\kappa\mu\nu}}
\newcommand{\Vlamunu}{V_{\lambda\mu\nu}}
\newcommand{\Vmunuro}{V_{\mu\nu\rho}}
\newcommand{\Vkalamunu}{V_{\kappa\lambda\mu\nu}}
\newcommand{\Fkala}{F_{\kappa\lambda}}
\newcommand{\Fkanu}{F_{\kappa\nu}}
\newcommand{\Flaka}{F_{\lambda\kappa}}
\newcommand{\Flamu}{F_{\lambda\mu}}
\newcommand{\Fmunu}{F_{\mu\nu}}
\newcommand{\Fnumu}{F_{\nu\mu}}
\newcommand{\Fnuka}{F_{\nu\kappa}}
\newcommand{\Fmuka}{F_{\mu\kappa}}
\newcommand{\Fkalamu}{F_{\kappa\lambda\mu}}
\newcommand{\Flamunu}{F_{\lambda\mu\nu}}
\newcommand{\Flanumu}{F_{\lambda\nu\mu}}
\newcommand{\Fkamula}{F_{\kappa\mu\lambda}}
\newcommand{\Fkanumu}{F_{\kappa\nu\mu}}
\newcommand{\Fmulaka}{F_{\mu\lambda\kappa}}
\newcommand{\Fmulanu}{F_{\mu\lambda\nu}}
\newcommand{\Fmunuka}{F_{\mu\nu\kappa}}
\newcommand{\Fkalamunu}{F_{\kappa\lambda\mu\nu}}
\newcommand{\Flakanumu}{F_{\lambda\kappa\nu\mu}}

\newcommand{\tvec}{\vec}
\newcommand{\action}{\mathscr{S}}
\newcommand{\pathdiff}[1]{\!\mathscr{D}#1\,}
\newcommand{\diff}[1]{\!\mathrm{d}#1\,}
\newcommand{\dottau}{\accentset{\boldsymbol\circ}}
\newcommand{\dott}{\accentset{\mbox{\large .}}}
\newcommand{\ddott}{\accentset{\mbox{\large ..}}}
\newcommand{\dd}[2][]{\frac{\mathrm{d} #1}{\mathrm{d} #2}}
\newcommand{\pdd}[2][]{\frac{\partial #1}{\partial #2}}
\newcommand{\atanh}{\operatorname{atanh}}
\newcommand{\sech}{\operatorname{sech}}
\newcommand{\keld}{\tilde{\gamma}}
\renewcommand{\Re}{\operatorname{Re}}
\renewcommand{\Im}{\operatorname{Im}}
\newcommand\numberthis{\addtocounter{equation}{1}\tag{\theequation}}

\newcommand{\ket}[1]{\left|#1\right>}
\newcommand{\bra}[1]{\left<#1\right|}
\newcommand{\braket}[2]{\left<#1|#2\right>}
\newcommand{\nn}{\nonumber\\}
\newcommand{\ul}{\underline}
\newcommand{\f}[1]{\mbox{\boldmath$#1$}}
\newcommand{\fk}[1]{\mbox{\boldmath$\scriptstyle#1$}}
\newcommand{\vau}{\mbox{\boldmath$v$}}
\newcommand{\na}{\mbox{\boldmath$\nabla$}}
\newcommand{\bea}{\begin{eqnarray}}
\newcommand{\ea}{\end{eqnarray}}
\newcommand{\eea}{\end{eqnarray}}
\newcommand{\ord}{\,{\cal O}}
\newcommand{\li}{\,\widehat{\cal L}}
\newcommand{\vc}[1]{\mathbf{#1}}
\newcommand{\sumint}[1]
{\begin{array}{c} \\
{{\textstyle\sum}\hspace{-1.1em}{\displaystyle\int}}\\
{\scriptstyle{#1}}
\end{array}}

\title{On the Heisenberg limit for detecting vacuum birefringence}

\author{N.~Ahmadiniaz} 
\affiliation{Helmholtz-Zentrum Dresden-Rossendorf, Bautzner Landstra\ss e 400, 01328 Dresden, Germany} 
\author{T.E.~Cowan}
\affiliation{Helmholtz-Zentrum Dresden-Rossendorf, Bautzner Landstra\ss e 400, 01328 Dresden, Germany} 
\affiliation{Institut f\"ur Kern-und Teilchenphysik, Technische Universit\"at Dresden, 01062 Dresden, Germany}
\author{R.~Sauerbrey}
\affiliation{Helmholtz-Zentrum Dresden-Rossendorf, Bautzner Landstra\ss e 400, 01328 Dresden, Germany} 
\affiliation{Institut f\"ur angewandte Physik, Technische Universit\"at Dresden, 01062 Dresden, Germany}
\author{U.~Schramm}
\affiliation{Helmholtz-Zentrum Dresden-Rossendorf, Bautzner Landstra\ss e 400, 01328 Dresden, Germany} 
\affiliation{Institut f\"ur Kern-und Teilchenphysik, Technische Universit\"at Dresden, 01062 Dresden, Germany}
\author{H.-P.~Schlenvoigt} 
\affiliation{Helmholtz-Zentrum Dresden-Rossendorf, Bautzner Landstra\ss e 400, 01328 Dresden, Germany} 
\author{R.~Sch\"utzhold}
\affiliation{Helmholtz-Zentrum Dresden-Rossendorf, Bautzner Landstra\ss e 400, 01328 Dresden, Germany} 
\affiliation{Institut f\"ur Theoretische Physik, Technische Universit\"at Dresden, 01062 Dresden, Germany}

\begin{abstract}
Quantum electrodynamics predicts the vacuum to behave as a non-linear medium, including effects such as 
birefringence. 
However, for experimentally available field strengths, this vacuum polarizability is extremely small and 
thus very hard to measure. 
In analogy to the Heisenberg limit in quantum metrology, we study the minimum requirements for such a 
detection in a given strong field (the pump field).
Using a laser pulse as the probe field, we find that its energy must exceed a certain threshold depending 
on the interaction time. 
However, a detection at that threshold, i.e., the Heisenberg limit, requires highly non-linear 
measurement schemes -- while for ordinary linear-optics schemes, the required energy 
(Poisson or shot noise limit) is much larger.
Finally, we discuss several currently considered experimental scenarios from this point of view. 
\end{abstract}

\date{\today}

\maketitle

\section{Introduction}\label{intro}

Classical electrodynamics is governed by the Maxwell equations, which are linear in the absence of sources.  
%
Thus electromagnetic waves in vacuum obey the superposition principle and do not interact \cite{jackson}. 
Quantum electrodynamics, on the other hand, predicts deviations from this behavior: 
Even the quantum vacuum should be polarizable and thus behave as a non-linear medium due 
to the coupling to the fermionic modes, see, e.g.,  
\cite{dirac28,euko35,eul36,euhe36,serber35,ueh35,sauter31,karneu51,weiss36,schw51}. 

However, since this polarizability is extremely weak for  available fields,
this fundamental prediction of quantum electrodynamics has not been experimentally 
verified yet for electromagnetic waves in vacuum (i.e., real photons).
Note that an analogous effect has been observed for the interaction of photons with 
the Coulomb field of atomic nuclei (referred to as Delbr\"uck scattering) 
\cite{del33,akhm98,milssch94,atlas}.

Partly motivated by the present-day and near-future experimental facilities aimed 
at the generation of ultra-strong electromagnetic fields, there have been several 
proposals and initiatives for verifying this fundamental prediction experimentally 
\cite{becker75,aleksander85,heinzel06,tommasini09,tommasini10,king12,gies13,gies14,
schle16,dipiazza-16,homma16,ataman17,ataman18,naka17,dinu141,dinu142,mignani17,capparelli17,shen},
see also \cite{bmw,pvlasfer,oval} and 
\cite{toll52,klenig64,baiber67,biabia70,dittrichbook,rikriz00,mark,dipiazza-17,karb15,karb16}.
%
%
In order to compare the different proposals and to sort them into a bigger picture, 
we address the general question of what are the minimum requirements for detecting 
the tiny polarizability or non-linearity of the quantum vacuum.   

\section{Euler-Heisenberg Lagrangian}\label{Euler-Heisenberg}

Let us start with a brief recapitulation of the basic principles. 
For slowly varying electromagnetic fields $\mathbf{E}$ and $\mathbf{B}$
well below the Schwinger critical field determined by the electron mass 
$m$ and the elementary charge $q$
\bea
E_{\rm crit}=\frac{m^2c^3}{\hbar q}\approx 1.3\times10^{18}\,\frac{\rm V}{\rm m}
\,,
\ea
i.e., the range we are interested in, the propagation can be described by the 
lowest-order Euler-Heisenberg Lagrangian density \cite{euko35,euhe36,sauter31,karneu51,dunne05}
\bea
\label{EHL}
\mathcal{L}
&=&
\frac{\varepsilon_0}{2}\left(\mathbf{E}^2-c^2\mathbf{B}^2\right)
\nn
&&
+
\xi\left[\left(\mathbf{E}^2-c^2\mathbf{B}^2\right)^2
+7c^2\left(\mathbf{E}\cdot\mathbf{B}\right)^2
\right]\,,
\ea
with the vacuum permittivity $\varepsilon_0$ and the pre-factor 
\bea
\xi=\frac{\hbar q^4}{360\pi^2 m^4c^7}=\varepsilon_0\,\frac{\alpha_{\rm QED}}{90\pi E_{\rm crit}^2}=\frac{2\alpha_{\rm QED}^2}{45m^4}
\,,
\ea
where $\alpha_{\rm QED}\approx1/137$ is the fine-structure constant. 
From now on, we shall employ natural units with 
\bea
\hbar=c=\varepsilon_0=1\,,
\ea
in order to simplify the expressions.

\subsection{Pump and probe field}

We consider the standard situation where we have a strong (but sub-critical) 
pump field ${\bf E}_0$ and ${\bf B}_0$ acting on the vacuum plus a weaker 
probe pulse ${\bf E}_1$ and ${\bf B}_1$ in order to detect the induced 
polarizability. 
The pump field is supposed to be a solution of the equations of motion 
stemming from~(\ref{EHL}) and we study the propagation of the probe field 
in this background. 
Then, inserting the split ${\bf E}={\bf E}_0+{\bf E}_1$ into ${\bf B}={\bf B}_0+{\bf B}_1$
into~\eqref{EHL} and linearizing the equations of motion in ${\bf E}_1$ and ${\bf B}_1$, 
we obtain the effective Lagrangian for the probe field 
\bea
\label{probe}
\mathcal{L}_1 
&=&
\half\left[
{\bf E}_1\cdot(\mathbb{1}+\delta\epsilon)\cdot{\bf E}_1-
{\bf B}_1\cdot(\mathbb{1}-\delta\mu)\cdot{\bf B}_1\right] 
\nn
&&
+{\bf E}_1\cdot\delta\Psi\cdot{\bf B}_1 
\,.
\ea
The polarizability of the vacuum is encoded in the change of the dielectric 
permittivity tensor $\delta\epsilon$ and the 
magnetic permeability tensor $\delta\mu$ 
as well as the symmetry breaking tensor $\delta\Psi$. 
These quantities depend on the strength of the pump field ${\bf E}_0$ and ${\bf B}_0$
(see the Appendix) and are suppressed as 
$\ord(\alpha_{\rm QED}[{\bf E}_0^2+{\bf B}_0^2]/E_{\rm crit}^2)$.
Since they are very small, we only keep their first order.  

Note that we consider the modifications in the propagation of the probe 
pulse induced by the pump field. 
Thus, we do not consider other non-linear QED effects such as 
photon splitting or four-wave mixing, see, e.g. 
\cite{adler71,akhm02,adler70,rozanov93,mckena63,varfo66,moulin99,dipiazza05,bernard00,lund06,
lundin06,gies1,gies2,king18}, which would 
correspond to linear or cubic powers of ${\bf E}_1$ and ${\bf B}_1$ 
in~\eqref{probe}.  
In the following, we shall focus on the probe field and consider the 
tensors $\delta\epsilon$, $\delta\mu$ and $\delta\Psi$ as externally 
given. 
Thus, we shall drop the sub-scripts for the probe field ${\bf E}_1$ and ${\bf B}_1$ 
from now on. 

\subsection{Interaction Hamiltonian}

In terms of the usual vector potential $\bf A$, the canonically conjugate momentum 
(which equals the dielectric displacement field $\bf D$) reads 
\bea
{\bf\Pi}={\bf D}=(\mathbb{1}+\delta\epsilon)\cdot {\bf E}+\delta\Psi\cdot {\bf B}
\,,
\ea
%
and thus the Hamiltonian density is given by 
\bea
\mathcal{H} 
&=&\half\left[
{\bf \Pi}\cdot(\mathbb{1}-\delta\epsilon)\cdot{\bf \Pi}+{\bf B}\cdot (\mathbb{1}-\delta\mu)\cdot{\bf B}
\right]
\nn
&&
-{\bf \Pi}\cdot\delta\Psi\cdot{\bf B}
\,.
\label{interactionH}
\ea
After splitting off the undisturbed vacuum contribution 
${\mathcal H}_0=[{\bf\Pi}^2+{\bf B}^2]/2$, the remaining part 
describes the interaction between the probe field and the polarizability 
$\delta\epsilon$, $\delta\mu$ and $\delta\Psi$ induced by the pump field 
\bea
\label{interaction}
\mathcal{H}_{\rm int}
=
-\half{\bf \Pi}\cdot\delta\epsilon\cdot{\bf \Pi}
-\half{\bf B}\cdot\delta\mu\cdot{\bf B}
-{\bf \Pi}\cdot\delta\Psi\cdot{\bf B}
\,.
\ea
%

\section{Heisenberg Limit}\label{Heisenberg-Limit}

As announced above, let us now study the question of which requirements the probe pulse has to fulfill in order 
to detect the vacuum polarizability $\delta\epsilon$, $\delta\mu$ and $\delta\Psi$.
According to the laws of quantum mechanics, this effect is only detectable if the quantum state $\ket{\psi}$ 
of the probe field after its interaction with the pump field $\hat{U}_{\rm int}\ket{\psi}$ deviates sufficiently 
from the quantum state $\ket{\psi}$ without this interaction. 
As one possible signature, the no-signal fidelity given by (see also \cite{sch18})
\bea
\langle\psi\vert\hat{U}_{\rm int}\vert\psi\rangle
=
\langle\psi\vert\mathcal{T}\exp\Big\{-i\int dt\;\hat{H}_{\rm int}(t)\Big\}\vert\psi\rangle\,,
\ea 
should sufficiently deviate from unity ($\mathcal{T}$ is the time ordering operator). 
Since the interaction Hamiltonian
\bea
\label{interaction-Hamiltonian}
\hat{H}_{\rm int}=\int d^3r\;\hat{\mathcal H}_{\rm int}\,,
\ea
is linear in the small tensors $\delta\epsilon$, $\delta\mu$ and $\delta\Psi$, let us 
apply first-order perturbation theory 
\bea
\langle\psi\vert\hat{U}_{\rm int}\vert\psi\rangle
=
1-i\int dt\;\langle\psi\vert\hat{H}_{\rm int}(t)\vert\psi\rangle
+\ord(\hat{H}_{\rm int}^2)
\,.
\ea 
We see that the lowest-order contribution corresponds to a phase shift \cite{sch18}
\bea
\label{phase-shift}
\varphi
=
-\int dt\;\langle\psi\vert\hat{H}_{\rm int}(t)\vert\psi\rangle
\,,
\ea 
which could be measured by interferometric means, for example 
(see below). 

\subsection{Classical fields}
\label{classical fields}
Let us estimate the maximum possible phase shift~\eqref{phase-shift} 
for a given probe pulse. 
First, we treat the probe pulse as a classical field, which should be 
a good approximation for laser pulses. 
Inserting Eqs.~\eqref{interaction-Hamiltonian} and \eqref{interaction} 
into \eqref{phase-shift}, we obtain space-time integrals over the terms 
${\bf \Pi}\cdot\delta\epsilon\cdot{\bf \Pi}/2$,  
${\bf B}\cdot\delta\mu\cdot{\bf B}/2$ and 
${\bf \Pi}\cdot\delta\Psi\cdot{\bf B}$. 
Since the tensor $\delta\epsilon$ is real and symmetric, we may diagonalize
it and obtain the bound 
\bea
{\bf \Pi}\cdot\delta\epsilon\cdot{\bf \Pi}\leq{\bf \Pi}^2||\delta\epsilon||
\,,
\ea
where the norm $||\delta\epsilon||$ is the maximum of the absolute values 
of the eigenvalues of $\delta\epsilon$. 
In complete analogy, we can bound the term ${\bf B}\cdot\delta\mu\cdot{\bf B}$
by the same norm $||\delta\mu||$ multiplied by ${\bf B}^2$.
%
Thus, we find 
\bea
\frac12\left[
{\bf \Pi}\cdot\delta\epsilon\cdot{\bf \Pi}
+{\bf B}\cdot\delta\mu\cdot{\bf B}
\right]
\leq
\mathcal{E}\,
{\rm max}
\left\{||\delta\epsilon||,||\delta\mu||\right\}\,,
\ea
at each space-time point, where $\mathcal{E}=({\bf \Pi}^2+{\bf B}^2)/2$ 
is the (lowest-order) energy density of the probe pulse. 

The remaining term ${\bf \Pi}\cdot\delta\Psi\cdot{\bf B}$ is a bit more 
complicated because the tensor $\delta\Psi$ is not symmetric in general. 
Thus, we employ the singular value decomposition 
\bea
\delta\Psi=\sum_I \sigma_I {\bf u}_I\otimes{\bf v}_I
\,,
\ea
with the non-negative singular values $\sigma_I$ and the two (left and right) 
orthonormal basis sets ${\bf u}_I$ and ${\bf v}_I$. 
Then, using 
$({\bf u}_I\cdot{\bf \Pi})({\bf v}_I\cdot{\bf B})\leq|{\bf \Pi}| |{\bf B}|\leq({\bf \Pi}^2+{\bf B}^2)/2$,
we arrive at 
\bea
\label{Heisenberg}
\varphi\leq T\,{\rm max}_{\bf r} 
\left\{
\sum_I \sigma_I
+
{\rm max}\left\{||\delta\epsilon||,||\delta\mu||\right\}
\right\}
\mathfrak E
\,,
\ea
where $T$ denotes the interaction time and $\mathfrak E$ the total energy 
of the probe pulse.
The spatial integral can be bounded from above by the maximum over all 
positions $\bf r$ since all the involved quantities, such as the energy 
density $\mathcal E$, are non-negative 
(for classical fields, quantum fields will be discussed in the next section). 

Turning the above argument around, we get a minimum energy $\mathfrak E$ 
of the probe pulse required for detecting the vacuum polarizability 
$\delta\epsilon$, $\delta\mu$ and $\delta\Psi$ in a given interaction time 
$T$ since the phase shift $\varphi$ should not be too small in order to 
achieve a measurable effect. 
Since the energy $\mathfrak E$ scales linearly with the number $N$ of probe 
photons, we refer to~\eqref{Heisenberg} as the Heisenberg limit. 

\subsection{Quantum fields}

In the previous section, we treated the probe pulse as a classical field in 
order to derive the bound~\eqref{Heisenberg}.
In the following, let us study whether an analogous bound can be established 
for quantum fields. 
As a crucial difference, expectation values such as $\langle\hat{\bf\Pi}^2\rangle$
or 
$\langle\hat{\bf\Pi}\cdot\delta\varepsilon\cdot\hat{\bf\Pi}\rangle$ are 
divergent and thus require renormalization. 
As usual, we achieve this by subtracting the vacuum expectation value 
\bea
\label{renormalization}
\langle\hat{\bf\Pi}^2\rangle_{\rm ren}
=
\bra{\psi}\hat{\bf\Pi}^2\ket{\psi}
-\bra{0}\hat{\bf\Pi}^2\ket{0}
\,.
\ea
Of course, this requires appropriate regularization.
Here, we use the normal mode decomposition. 
To this end, we introduce a complete set of orthonormal 
\bea
\int d^3r\;{\bf f}_I\cdot{\bf f}_J=\delta_{IJ}\,,
\ea
and transversal $\nabla\cdot{\bf f}_I=0$ basis functions ${\bf f}_I({\bf r})$
and expand the field operates into this basis set 
\bea
\hat{\bf\Pi}(t,{\bf r})=\sum_I\hat p_I(t){\bf f}_I({\bf r})
\,.
\ea
Inserting this normal mode decomposition, we find 
\bea
\int d^3r\;\hat{\bf\Pi}\cdot\delta\varepsilon\cdot\hat{\bf\Pi}
&=&
\sum_{IJ}\hat p_I\hat p_J 
\int d^3r\;{\bf f}_I\cdot\delta\varepsilon\cdot{\bf f}_I
\nn
&=&
\sum_{IJ}\hat p_I\hat p_J\,M_{IJ} 
\,.
\ea
After diagonalizing this real and symmetric matrix $M_{IJ}$ via the 
orthogonal matrix $O_{IJ}$, we may introduce a new set of basis 
functions via ${\bf F}_I=\sum_J O_{IJ}{\bf f}_J$ and expand the 
field operator in this new set 
$\hat{\bf\Pi}(t,{\bf r})=\sum_I\hat P_I(t){\bf F}_I({\bf r})$ 
which gives the simplified expression 
\bea
\int d^3r\;\hat{\bf\Pi}\cdot\delta\varepsilon\cdot\hat{\bf\Pi}
=
\sum_{I}\lambda_I\hat P_I^2 
\,,
\ea
where $\lambda_I$ are the eigenvalues of the matrix $M_{IJ}$.
Since the $\hat P_I^2$ are positive operators, we may even derive 
a (formal) bound on the operator level 
\bea
\label{operator-bound}
\int d^3r\;\hat{\bf\Pi}\cdot\delta\varepsilon\cdot\hat{\bf\Pi}
\leq
||M||\sum_{I}\hat P_I^2 
=
||M||\sum_{I}\hat p_I^2 
\,,
\ea
where $||M||={\rm max}_I|\lambda_I|$ is the norm of the matrix $M_{IJ}$ 
in analogy to the previous section.
It can be estimated by the maximum ``expectation value'' 
$\int d^3r\;{\bf f}\cdot\delta\varepsilon\cdot{\bf f}$ 
for normalized functions $\bf f$ and thus agrees with 
${\rm max}_{\bf r}||\delta\varepsilon||$.

The above operator-valued bound~\eqref{operator-bound} seems to be 
the proper quantum generalization of the Heisenberg limit~\eqref{Heisenberg}
since the sums $\frac12\sum_{I}\hat p_I^2$ and 
$\frac12\sum_{I}\hat P_I^2$ 
can be bounded by the total energy of the probe pulse which reads 
$\frac12\sum_{I}(\hat p_I^2+\Omega_I^2\hat q_I^2)$ in terms of the 
eigenmodes $I$ with the eigenfrequencies $\Omega_I$. 
However, this bound is of limited use since the expectation value 
diverges due to the infinite zero-point energy (as mentioned above). 
After subtracting this zero-point energy~\eqref{renormalization}, 
we cannot deduce the inequality $\langle\hat p_I^2\rangle_{\rm ren}
\leq\langle\hat p_I^2\rangle_{\rm ren}+\Omega_I^2\langle\hat q_I^2\rangle_{\rm ren}$ 
anymore 
since the renormalized expectation values 
$\langle\hat p_I^2\rangle_{\rm ren}$ and 
$\langle\hat q_I^2\rangle_{\rm ren}$ can become negative.
For example, in a squeezed state $\ket{0}\to\ket{r}$, 
we may increase the momentum variance 
$\langle\hat p_I^2\rangle\to\exp\{+r\}\langle\hat p_I^2\rangle$
while decreasing the position variance 
$\langle\hat q_I^2\rangle\to\exp\{-r\}\langle\hat q_I^2\rangle$, 
such that it is below its ground-state value 
$\langle\hat q_I^2\rangle<\bra{0}\hat q_I^2\ket{0}$ which means that 
$\langle\hat q_I^2\rangle_{\rm ren}$ becomes negative. 

Thus, if we naively replace the classical energy $\mathfrak E$ in the 
Heisenberg limit~\eqref{Heisenberg} by the renormalized expectation 
value ${\mathfrak E}_{\rm ren}$ for quantum fields, it would be possible 
to violate this bound by squeezing many modes just a little bit $r\ll1$
such that their $\langle\hat p_I^2\rangle_{\rm ren}\sim r$ 
increase while the growth of the energy ${\mathfrak E}_{\rm ren}\sim r^2$ 
is suppressed. 
One could suspect that this enhancement would be compensated by the 
other terms such as $\hat{\bf B}\cdot\delta\mu\cdot\hat{\bf B}$
which contains $\langle\hat q_I^2\rangle_{\rm ren}$ but this is not 
the case since different modes contribute differently to these terms. 
Thus, one could squeeze those modes where the first term 
$\hat{\bf\Pi}\cdot\delta\varepsilon\cdot\hat{\bf\Pi}$
dominates in one way 
$\langle\hat p_I^2\rangle\to\exp\{+r\}\langle\hat p_I^2\rangle$ 
and the other modes where the second term 
$\hat{\bf B}\cdot\delta\mu\cdot\hat{\bf B}$
dominates in the opposite way 
$\langle\hat p_I^2\rangle\to\exp\{-r\}\langle\hat p_I^2\rangle$. 
For simplicity, we have omitted the third term $\propto\delta\Psi$ 
since it has yet another mode structure, see also Sec.~\ref{PVLAS}. 

In summary, the divergent zero-point energy invalidates a bound like~\eqref{Heisenberg} 
for quantum electrodynamics.
To obtain a generalized bound, one would have to limit the number of involved 
modes $I$ as well as their eigenfrequencies $\Omega_I$, which is difficult
\footnote{
Note that this problem arises for relativistic quantum fields.
For non-relativistic quantum fields, such as the Schr\"odinger field $\hat\Psi$ describing the 
atoms in a Bose-Einstein condensate, it is possible to obtain bounds similar 
to~\eqref{Heisenberg}.  
For example, the interaction $\hat{\mathcal H}_{\rm int}$ 
of a Bose-Einstein condensate with a gravitational wave 
(or similar disturbances) can be split into the contribution 
$\hat\Psi^\dagger\delta V\hat\Psi$
stemming from the effective change $\delta V$ of the trap potential 
plus the term 
$(\nabla\hat\Psi^\dagger)\cdot\delta g\cdot(\nabla\hat\Psi)/2m$
induced by the variation $\delta g$ of the spatial metric 
describing the gravitational wave \cite{sch18}. 
%
%
%
Due to the absence of a zero-point energy, we may derive a bound in analogy 
to~\eqref{Heisenberg} in terms of the total number of atoms $N$ and the 
total kinetic energy ${\mathfrak E}_{\rm kin}$.}.  

\section{Comparison to Poisson Limit}\label{Shot-Noise-Limit}

Coming back to the Heisenberg limit~\eqref{Heisenberg}, one might object that a global 
phase $\varphi$ cannot be measured. 
While this is correct in principle, this objection could be circumvented by considering a 
scenario involving a quantum superposition of two paths of the probe pulse, one interacting 
with the pump field and the other one not. 
This state corresponds to a NOON state \cite{kok02,pryde03,cable07}
\bea
\label{NOON}
\ket{\psi}_{\rm NOON}=\frac{\ket{N,0}+\ket{0,N}}{\sqrt{2}}
\,,
\ea
where either all $N$ probe photons take the one path $\ket{N,0}$ or all $N$ probe photons 
take the other path $\ket{0,N}$. 
Note that this is a highly non-classical state, in analogy to the Greenberger-Horn-Zeilinger 
(GHZ) state \cite{GHZ1,GHZ2}.  
After interaction with the pump field (in one path only), this state evolves into 
$(\ket{N,0}+e^{i\varphi}\ket{0,N})/\sqrt{2}$ which becomes orthogonal to the initial 
state~\eqref{NOON} for $\varphi=\pi$. 
Note, however, that both, preparing the initial state~\eqref{NOON} as well as measuring 
the final state $(\ket{N,0}+e^{i\varphi}\ket{0,N})/\sqrt{2}$ would require effectively 
$N$-photon interactions, i.e., a highly non-linear optics scheme. 

In a typical linear optics set-up the (coherent) state of a laser is described 
by the factorizing state 
\bea
\label{classical}
\ket{\psi}_{\rm laser}=\bigotimes_{\ell=1}^N\frac{\ket{1,0}_\ell+\ket{0,1}_\ell}{\sqrt{2}}
\,,
\ea
where each photon $\ell$ individually either takes the one path $\ket{1,0}$ 
or the other path $\ket{0,1}$. 
In this case, one would obtain a Poisson distribution of the photon numbers 
in the output channel and thus the accuracy scales with $1/\sqrt{N}$ instead of $1/N$, 
which is the well-known classical Poisson (shot-noise) limit. 

Let us illustrate this distinction in terms of the scaling of the phase with photon number $N$. 
According to the Heisenberg limit~\eqref{Heisenberg}, we find 
\bea
\Delta\varphi_N=N\Delta\varphi_1
\,,
\ea
where $\Delta\varphi_1$ is the phase shift experienced by a single 
photon with frequency $\omega$ 
\bea
\Delta\varphi_1
=
\omega T\,{\rm max}_{\bf r} 
\left\{
\sum_I \sigma_I
+
{\rm max}\left\{||\delta\epsilon||,||\delta\mu||\right\}
\right\}
\,. 
\ea
Since $\Delta\varphi_N$ must be of order unity to obtain a measurable 
detection probability, we get the well-known Heisenberg scaling 
$\Delta\varphi_1\sim1/N$. 

In contrast, the Poisson distribution of the photon numbers in the 
classical (i.e., coherent) state~\eqref{classical} results in a relative 
accuracy of $1/\sqrt{N}$ which yields the well-known Poisson limit 
$\Delta\varphi_1\sim1/\sqrt{N}$, see \cite{xiao87}. 

\section{Experimental Scenarios}\label{Experimental}

\subsection{Static magnetic pump field}\label{PVLAS}

There are several running or planned experiments where the pump field is 
an approximately constant magnetic field of 
a few Tesla, see, e.g., \cite{pvlasfer,pvlas}.
For a purely magnetic field, the symmetry-breaking term $\delta\Psi$ vanishes 
(see the Appendix).
The maximum eigenvalues of the remaining terms $\delta\varepsilon$ and $\delta\mu$ 
are given by  $10\xi B_0^2$ and $12\xi B_0^2$, respectively, which are then of order $10^{-22}$.  
Thus, the accuracy requirements are roughly comparable to those for the 
detection of gravitational waves at LIGO \cite{ligo16}. 

As in LIGO, the signal can be amplified by having the probe photons bounce back and forth 
many times (in a cavity, for example), which facilitates a large integration time $T$.
%
Assuming an optimized
cavity finesse of order $10^6$ and length scales of order meter, 
we get an integration time of order 
$\omega T=\ord(10^{12})$ periods for optical or near-optical photons.
Again in analogy to LIGO, the remaining orders of magnitude should be compensated 
by a sufficiently large number of probe photons.
Using the Heisenberg limit~\eqref{Heisenberg}, we would get $N=\ord(10^{10})$ 
which is a comparably low number. 
However, as explained above, this detection scheme would require an effective 
$N$-photon interaction involving this number of photons, which is currently 
out of reach. 

With laser fields and linear optics schemes, we can only reach the Poisson 
limit, 
which gives $N=\ord(10^{20})$ corresponding to a probe pulse energy $\mathfrak E$ 
in the Joule regime. 
This shows that such an experiment is not impossible with present-day-technology 
but still quite challenging. 

Note that the actual limit is even a bit larger because these experiments 
typically do not measure the polarizabilities $\delta\varepsilon$ and 
$\delta\mu$ directly, but only the induced rotation of polarization -- 
which measured their difference in the different directions. 
Otherwise, the rotation of polarization is very similar to an interferometric
set-up, where the two arms correspond to the two polarizations.

\subsection{Optical pump and XFEL probe}\label{XFEL} 

Another popular scheme envisions a strongly focused ultra-strong laser pulse 
(again in the optical or near-optical regime) where intensities of order 
 $10^{22}~\rm W/cm^2$ or more 
should be reachable with present-day or 
near-future technology see, e.g., \cite{xfel,eli,corels}.
This corresponds to electric fields above $10^{14}~\rm V/m$
which generate polarizabilities $\delta\epsilon$, $\delta\mu$ and $\delta\Psi$ 
of order $10^{-11}$.
%
This illustrates a major advantage in comparison to the static set-up 
in the previous section~\ref{PVLAS}, as the pump field is much stronger 
in a laser focus. 
As a drawback, the interaction time $T$ is limited to the pump pulse length 
of a few (say ten) optical cycles. 

However, for a probe pulse generated by an x-ray free electron laser (XFEL) 
with photon energies in the 10~keV range, this corresponds to $\ord(10^5)$
XFEL cycles, see, e.g., \cite{schle16}.  
The Heisenberg limit~\eqref{Heisenberg} then gives 
$N=\ord(10^6)$ 
photons, i.e., an energy of 
 $\ord(10^{10}~\rm eV)$ or $\ord(10^{-9}~\rm J)$.
%
Again, as an $N$-photon interaction with these numbers seems unrealistic, 
the Poisson 
limit yields $N=\ord(10^{12})$ 
photons, corresponding to an energy of 
 $\ord(10^{16}~\rm eV)$ or $\ord(10^{-3}~\rm J)$.
%
As before, this shows that the detection is challenging but not completely 
out of reach.  

In analogy to the previous section~\ref{PVLAS}, the envisioned scheme 
is based on the rotation of the polarization which offers experimental
advantages in comparison to an interferometric set-up with x-rays, 
but decreases the signal a bit. 
Note that, with $N=\ord(10^{11})$ 
photons in an initially 
polarized probe beam 
(see also \cite{xfel-parameters-1,xfel-parameters-2,xfel-parameters-3,marx,bernhardt}),  
the signal 
 may consist of a single photon with flipped polarization after several runs.
%
This necessitates a careful study of potentially competing effects 
in order to distinguish the signal from the background. 

\subsection{Optical pump and optical probe}  

In contrast to the scenario described above, one could also consider 
an optical (or near-optical) probe pulse, see also \cite{ataman18}. 
Using the same parameters for the pump pulse, the Heisenberg 
limit~\eqref{Heisenberg} would yield the same energy $\mathfrak E$ 
as in the previous section~\ref{XFEL}.  
However, the probe pulse would now contain $N=\ord(10^{10})$ 
photons because the interaction time corresponds to a few optical cycles only. 
The Poisson 
limit then yields $N=\ord(10^{20})$ 
corresponding to an energy of $\ord(10^{20}~\rm eV)$ or $\ord(10~\rm J)$.
%
The fact that this is of the same order as the 
pump pulse itself shows the challenges of this detection scheme.

On the other hand, for this all-optical scheme, 
it is not necessary to have the optical laser close to an XFEL. 
Thus, it might be possible to reach even higher intensities in the 
$10^{23}~\rm W/cm^2$ regime, which reduces the requirements on the 
probe pulse to $N=\ord(10^{18})$ photons, i.e., an energy of 
$\ord(10^{18}~\rm eV)$ or $\ord(10^{-1}~\rm J)$.
As one possibility, one could use a dual-beam facility 
(see, e.g., \cite{draco}) or spit off a small part of the pump pulse 
before focusing and use it as probe pulse.
This could help ensuring the necessary temporal overlap between pump 
and probe pulse, which can pose a challenge for the scheme described 
in the previous section~\ref{XFEL}. 
Still, performing interference experiments with such $\ord(10^{-1}~\rm J)$ 
pulses containing $N=\ord(10^{18})$ photons is highly non-trivial.
Again in analogy to LIGO, it might be advantageous to operate the 
interferometer not exactly at the dark spot, but close to it -- 
corresponding to a small phase mismatch $\varphi_0$ between the two 
arms. 
For example, using a phase mismatch of $\varphi_0=10^{-3}$ 
(corresponding to a precision of placing the mirrors in the 
nanometer regime), the output at the darker port would contain 
$\ord(10^{12})$ photons. 
A single-photon phase shift of $\Delta\varphi_1=\ord(10^{-9})$
would then generate a signal of $\ord(10^{6})$ photons difference 
on top of the background of $\ord(10^{12})$ photons.  
Measuring such a large photon number with a relative accuracy of 
$\ord(10^{-6})$ is challenging and probably requires an advanced 
detector with many mega-pixels.  

On the other hand, this set-up could also offer an advantage as the 
signal would now contain many $\ord(10^{6})$ coherent photons, 
which could help distinguishing it from the background, 
especially from incoherent noise. 

\subsection{Angular dependence}

Especially for the latter all-optical scenario, it is probably unrealistic 
to assume a head-on collision between pump and probe pulse.
Thus, let us estimate the angular dependence of the phase shift. 
For simplicity, we model pump and probe pulse as plane waves (as a first step). 
Then, their relative directions can be described in terms of the 
three Euler angles $\psi$, $\theta$, and $\phi$. 
Without loss of generality, the pump field is supposed to propagate in 
$z$ direction with ${\bf E}_0$ and ${\bf B}_0$ pointing in $x$ and $y$ 
directions, respectively. 
After starting with the same orientation, the probe field directions 
are obtained by three rotations:
First, a rotation around the $z$ axis with the angle $\phi$, 
second, a rotation around the new $x$ axis with the angle $\theta$, 
followed by a third rotation around the new $z$ axis with the angle $\psi$. 

Then, the interaction Hamiltonian reads 
\bea 
{\mathcal H}_{\rm int}=-2\xi {\bf E}_0^2{\bf E}_1^2
\left(11-3\sin[2\psi-2\phi]
\right)\sin^4\frac{\theta}{2} 
\,.
\ea
For the co-propagating case $\theta=0$, it vanishes identically 
(as is well known). 
In this case, the angles $\psi$ and $\phi$ just rotate the polarization 
and one can transform into a co-moving Lorentz frame where all fields 
${\bf E}_0$ and ${\bf B}_0$ as well as ${\bf E}_1$ and ${\bf B}_1$
become arbitrarily small. 
The maximum is obtained in the counter-propagating case $\theta=\pi$
where such a Lorentz boost diminishing all fields is not possible. 
In this limiting case, the angles $\psi$ and $\phi$ again just rotate 
the polarization and the maximum signal is obtained by $\psi-\phi=-\pi/4$. 

Apart from these well-known limiting cases, we see that this condition 
$\psi-\phi=-\pi/4$ does also give the maximum signal for arbitrary given 
$\theta$, which may be unavoidable due to experimental constraints. 
For small deviations $\Delta\theta=\theta-\pi$ from the optimal 
counter-propagating case $\theta=\pi$, we find 
\bea 
{\mathcal H}_{\rm int}=-28\xi{\bf E}_0^2{\bf E}_1^2
\left(1-\frac{[\Delta\theta]^2}{2}+\ord([\Delta\theta]^4)
\right)
\,.
\ea
%

\section{Conclusions}\label{Conclusions}

We study the general requirements for detecting the weak vacuum 
polarizability~\eqref{probe} predicted by quantum electrodynamics.  
We find that the lowest-order effect is a phase shift~\eqref{phase-shift}
which could, at least in principle, be detected by interferometric means.
Approximating the probe pulse by a classical field, we obtain an upper 
bound~\eqref{Heisenberg} for the phase shift depending on the interaction
time $T$ and the total energy $\mathfrak E$ of the probe pulse. 

Since this phase shift must be of order unity for a measurable 
detection probability, this inequality does also give the Heisenberg 
limit (note that $\mathfrak{E}T$ equals the number of photons $N$ times 
the number of periods $\omega T$).  
However, such a measurement would require a highly non-linear optics scheme,
which is out of reach for realistic parameters. 
For a linear-optics scheme, we recover the well-known Poisson
(shot-noise) limit. 

Going beyond the classical field approximation, the failure of proving a bound 
as~\eqref{Heisenberg} for quantum fields hints at the interesting (theoretical) 
possibility to reach an even higher accuracy by exploiting the zero-point fluctuations. 
One option could be to squeeze many field modes a little bit such that their 
quadratures are modified (in order to increase the sensitivity) 
while the total energy expectation value does not change significantly. 
For optical frequencies, such a squeezing could be achieved in non-linear crystals 
in analogy to parametric down conversion, while for XFEL frequencies, a corresponding 
undulator set-up could serve the same goal, see also \cite{table-top}. 
However, apart from preparing this squeezed state initially, 
reading out the final state poses grand experimental challenges. 


Altogether, we obtain three regimes: the linear-optics regime corresponding 
to the Poisson (shot-noise) limit, the Heisenberg limit~\eqref{Heisenberg} for 
(locally) classical fields, and a regime beyond that limit for quantum 
fields. 

As a demonstration, we apply these limits to three experimental scenarios,
which offer different advantages (e.g., control of polarization for XFEL fields)
and disadvantages. 
For all cases, we find that the detection of the vacuum polarizability is quite 
challenging but not completely out of reach.  
Note that, apart from the verification aspect, the vacuum polarizability 
could also provide a clean way to measure the peak intensity of the laser, 
which is a highly non-trivial task. 

\acknowledgments 

N.A.~gratefully acknowledges helpful discussions with 
Felix Karbstein and Christian Kohlf\"urst. 
R.S.~acknowledges support from Deutsche Forschungsgemeinschaft 
(DFG, German Research Foundation, grant 278162697 -- SFB 1242).

\appendix

\begin{widetext}

\section{Vacuum permittivity and permeability tensors} \label{app2}

After inserting the split ${\bf E}={\bf E}_0+{\bf E}_1$ into ${\bf B}={\bf B}_0+{\bf B}_1$
into~\eqref{EHL} and keeping the quadratic terms only, we find the effective Lagrangian
for the probe field 
\bear
\mathcal{L}_1
=
\xi\Big\{
4({\bf E}_1\cdot{\bf E}_0-{\bf B}_1\cdot{\bf B}_0)^2+2({\bf E}_0^2-{\bf B}_0^2)({\bf E}_1^2-{\bf B}_1^2)
+7({\bf E}_1\cdot{\bf B}_0+{\bf B}_1\cdot{\bf E}_0)^2+14({\bf E}_0\cdot{\bf B}_0)({\bf E}_1\cdot{\bf B}_1)
\Big\}
\,.
\ear 
This effective Lagrangian provides information about the vacuum polarization tensors 
($\delta\epsilon$, $\delta\mu$ and $\delta\psi$) which can be extracted after comparing with \eqref{probe}
\bear
\mathcal{L}_1
&=&
\half\,{\bf E}_1\cdot \delta\epsilon\cdot{\bf E}_1+\half\,{\bf B}_1\cdot\delta\mu\cdot{\bf B}_1+
{\bf E}_1\cdot\delta\Psi\cdot{\bf B}_1
\,.
\label{efl}
\ear
Therefore these tensors are given in terms of the electric and magnetic components of the pump pulse 
\bear
\delta\epsilon_{ij}=2\xi\Big(2({\bf E}_{0}^2-{\bf B}_0^2)\delta_{ij}+4E_{0i}E_{0j}+7B_{0i}B_{0j}\Big)
\ear
which is a symmetric tensor, $\delta\mu$ is obtained from $\delta\epsilon$ by interchanging 
$\bf E$ and $\bf B$
\bear
\delta\mu=(\delta\epsilon)\Big\{E_i\leftrightarrow B_i, {\bf E}\leftrightarrow{\bf B}\Big\}
\,,
\ear
and finally the symmetry-breaking tensor
\bear
\delta\Psi_{ij}=2\xi\Big(7{\bf E}_0\cdot {\bf B}_0\delta_{ij}+7E_{0j}B_{0i}-4E_{0i}B_{0j}\Big)
\ear
  
\end{widetext}

%
As it has been already discussed in Section~\ref{Heisenberg-Limit}, 
we need to compute the eigenvalues of $\delta\epsilon$ and $\delta\mu$ 
and the singular values of $\delta\psi$ since the latter is in general an asymmetric tensor. 
It is straightforward to obtain the eigenvalues $\lambda_{1,2,3}$ of the symmetric tensors.  
For a general pump field, we obtain for $\delta\epsilon$ 
\bear
\lambda_{1,2}&=&\xi(3{\bf B}_0^2+8{\bf E}_0^2)\non
&&\pm\xi\sqrt{49{\bf B}_0^4+16{\bf E}_0^4-56{\bf B}_0^2{\bf E}_0^2+112({\bf E}_0\cdot {\bf B}_0)^2}\non
\lambda_3&=&4\xi({\bf E}_0^2-{\bf B}_0^2)\,.
\label{epsilon-lambda}
\ear 
In the limit of constant crossed fields they simplify to 
\bear
\vert\lambda_{1}\vert&=&14\xi{\bf E}_0^2\non
\vert\lambda_2\vert&=&8\xi{\bf E}_0^2\non
\lambda_3&=&0\,.
\label{epsilon-lambda-p}
\ear
Then the spectral representation of $\delta\epsilon$ is 
\bear
\delta\epsilon=\sum_{I=1}^3 \lambda_I {\bf v}_I \otimes {\bf v}_I\,,
\ear
where ${\bf v}_I$ are the eigenvectors of $\delta\epsilon$ 
(after a siutable rotation of the coordinate system) 
$$
{\bf v}_1=(1,0,0)
\,,\quad 
{\bf v}_2=(0,1,0)
\,,\quad
{\bf v}_3=(0,0,1) \,,
$$
and for $\delta\mu$ (with its eigenvalues are denoted by $\Lambda_I$) 
the eigenvalues are simply obtained from the ones for $\delta\epsilon$ by the following replacements 
\bear
\Lambda_1&=&\lambda_1(E_i\leftrightarrow B_i)\non
\Lambda_2&=&\lambda_2(E_i\leftrightarrow B_i)\non
\Lambda_3&=&\lambda_3(E_i\leftrightarrow B_i)\,.
\ear
In the limit of plane-wave background (or constant crossed fields), the two sets of eigenvalues for 
$\delta\epsilon$ and $\delta\mu$ become equivalent -- so to obtain the upper bound, 
it suffices to consider the maximum value of one of them. 
Bounding the terms in the Hamiltonian with $\delta\epsilon$ and $\delta\mu$ 
is simple since they have real eigenvalues, therefore one can consider the maximum value 
of their eigenvalues as discussed in Section~\ref{classical fields}. 
The most nontrivial term is the one with $\delta\psi$ since it has no symmetry 
in general then a direct eigenvalue computation may lead to imaginary values 
which correspond to non-orthogonal set of eigenvectors. 
Therefore one can rely on singular value decomposition which are defined 
based on the following theorem \cite{golb}:\newline
 
If $A$ is a real $m\times n$ matrix then there exist two orthogonal matrices 
$U=[{\bf u}_1,\cdots,{\bf u}_m]\in \mathbb{R}^{m\times m}$
and
$V=[{\bf v}_1,\cdots,{\bf v}_n]\in \mathbb{R}^{n\times n}$
such that 
\bear
U^TAV={\rm diag}[\sigma_1,\cdots,\sigma_p]\in \mathbb{R}^{m\times n}
\,,
\ear
where $p={\rm min}\{m,n\}$ and 
\bear
\sigma_1\ge\sigma_2\ge\cdots\ge\sigma_p\ge0
\,.
\ear
In other words the {\it singular values} $\sigma_1,\cdots,\sigma_p$ of a $m\times n$ matrix $A$ 
are the positive square roots, $\sigma_I=\sqrt{\lambda_I}>0$, of the nonzero eigenvalues of the 
associated {\it Gram} matrix $K=A^T A$. 
The corresponding eigenvectors of $K$ are known as the {\it singular vectors} of $A$ 
(note that for $m\ne n$ or rectangular matrices there is no eigenvalues in its general 
definition and that is why one finds the singular values). 
\newline 

This theorem lead to the following singular decomposition for a non-symmetric matrix 
$A\in \mathbb{R}^{n\times n}$
\bear
A=\sum_{I=1}^n\sigma_I {\bf u}_I \otimes {\bf v}_I\,,
\ear
and from here 
\bear
AA^T=\sum_{I=1}^n\sigma_I^2 {\bf u}_I \otimes {\bf u}_I\,,  
\ear
in which $\sigma_I$ are the singular values and the left and right singular vectors 
$u_I$ and $v_I$ for $I=1,2,\cdots,n$ respectively. 
Applying this theorem to $\delta\psi$ we get  
\bear
\sigma_{1,2}^2
&=&
2\xi^2\Big[65{\bf E}_0^2{\bf B}_0^2+84({\bf E}_0\cdot{\bf B}_0)^2\Big]
\non
&&
\pm 6\xi^2\sqrt{121{\bf B}_0^4{\bf E}_0^4+168{\bf E}_0^2{\bf B}_0^2({\bf E}_0\cdot{\bf B}_0)^2}\non
\sigma_3^2&=&196\xi^2({\bf E}_0\cdot {\bf B}_0)^2
\,.
\ear
For a constant crossed field they give  
\bear
\vert\sigma_{1}\vert&=&14\xi{\bf E}_0^2\non
\vert\sigma_2\vert&=&8\xi{\bf E}_0^2\non
\sigma_3&=&0
\,.
\ear 
After having the eigenvalues of $\delta\epsilon$ and $\delta\mu$ as well as the singular values of 
$\delta\psi$ one can easily compute the phase given in (\ref{Heisenberg}).

\section{The rotation matrix}
\label{euler}

Let us consider two different frames for the pump and probe pulses in which the former 
is fixed to be denoted by $xyz$ and the latter $XYZ$. 
We need three Euler angles to rotate $XYZ$ \cite{goldstein}.  
The sequence of the rotations are the following: 
$XYZ$ rotates by an angle $\phi$ about the $Z$-axis to obtain 
$\xi\eta\zeta$ with corresponding rotation matrix $\mathcal{D}$. 
For the second rotation, $\xi\eta\zeta$ is rotated about the 
$\xi$-axes by an angle $\theta$ to obtain new axes called 
$\xi'\eta'\zeta'$ with rotation matrix $\mathcal{C}$.  
Finally in the last step the later is rotated by an angle 
$\psi$ about $\zeta'$ to obtain $xyz$ with rotation matrix $\mathcal{B}$. 
The three successive rotations lead to a transformation matrix $\mathcal{A}$
\begin{widetext}
\bear
\mathcal{A}=
\mathcal{B}\,\mathcal{C}\mathcal{D}
=\begin{pmatrix} 
  \cos\psi\cos\phi-\cos\theta\sin\phi\sin\psi\,&\,\cos\psi\sin\phi+\cos\theta\cos\phi\sin\psi\,&\,\sin\psi\sin\theta\\
  \\
-\sin\psi\cos\phi-\cos\theta\sin\phi\cos\psi\,~&~-\sin\psi\sin\phi+\cos\theta\cos\phi\cos\psi~&~\sin \theta   \cos\psi\\
\\
\sin\theta\sin\phi~&~-\sin\theta\cos\phi~&~ \cos\theta\\
\\
\end{pmatrix}\,.
\ear
\end{widetext}
Therefore we have the following equation 
\bear
{\bf X}=\mathcal{A}\,{\bf x}\,,
\ear
where ${\bf x}=(x,y,z)$ and ${\bf X}=(X,Y,Z)$. 
To obtain a general form for the interaction Hamiltonian we need the 
probe electric and magnetic fields $({\bf E}_\omega, {\bf B}_{\omega})$. 
For a general probe field we have the following transformation 
\bear
{\bf E}_\omega^{\bf x}&=&\mathcal{A}\,{\bf E}_\omega^{\bf X}
\non
{\bf B}_\omega^{\bf x}&=&\mathcal{A}\,{\bf B}_\omega^{\bf X}\,,
\ear
if we consider a constant crossed background for the probe pulse 
in which the propagation direction lies again in $Z$ and ${\bf E}_\omega$ 
and ${\bf B}_\omega$ in $X$ and $Y$ directions accordingly then the interaction 
Hamiltonian defined in (\ref{interactionH}) with the polarization tensors obtained in 
Appendix~\ref{app2} one arrives at
\bear
\mathcal{H}_{\rm int}
&=&
-2\xi{\bf E}_0^2{\bf E}_\omega^2\Big(11-3\sin(2\psi-2\phi)\Big)\,\sin^4\left(\frac{\theta}{2}\right)\,,\non
\label{h}
\ear
which has a maximum at $(\theta=\pi,\psi-\phi=-\pi/4)$.

\end{document}